\documentclass[preprint,aps,pre]{revtex4}
\usepackage[dvips]{graphicx}

\begin{document}

\title{Nonlinear curvature elasticity of nematic liquid crystals}

\author{I. Lelidis$^{1}$ and G. Barbero$^{2,3}$ }
\affiliation{$^1$ Faculty of Physics, National and Kapodistrian University of Athens, \\
Panepistimiopolis, 15784 Zografos, Athens, Greece\\ $^2$Dipartimento di Scienza Applicata del Politecnico di Torino,\\
Corso Duca degli Abruzzi 24, 10129 Torino, Italia.\\
$^3$National Research Nuclear University MEPhI (Moscow Engineering Physics Institute), Kashirskoye shosse 31, 115409 Moscow, Russian Federation.
}
\date{\today}

\begin{abstract}
	The nonlinear elastic properties of nematic liquid crystals have acquired new interest with the recent experimental observation of bulk modulated nematic phases which are composed by achiral molecules.
We extend the Oseen-Zocher-Frank's elastic theory for nematic liquid crystals by including gradients of the nematic strain tensor in the elastic deformation energy. The invariants of the elastic tensor fields, up to the fourth order in the nematic director spatial derivatives, are calculated. An alternative approach that consists in the extension of the linear elastic energy to higher powers of the nematic strain tensor, as in classical elasticity of solids, is also developed. The twist-bend nematic modulated phase is investigated by both approaches and the results are critically compared. The conical angle of the twist-bend phase is calculated as function of the elastic constants. Surface-like effects are considered. Finally, we demonstrate that a splay-bend nematic phase with small oscillations of the nematic director around an axis is prohibited.
\end{abstract}

\pacs{}
\maketitle

	%%%%%%%%%%%%%%%%%%%%%%%%%%%%%%%%%%%%%%%%%%%%%%%%%%%%%
\section{Introduction}

%%%%%%%%%%%%%%%%%%%%%%%%%%%%%%%%%%%%%%%%%%%%%%%%%%%%%%%%%%

	The elastic distortion energy, $f$, of a uniaxial nematic in terms of the nematic director, $\mathbf{n}(\mathbf{r})$, field was developed by Oseen, Zocher, and Frank (OZF) \cite{oseen,zocher,frank}. Frank critically reformulated the elastic theory as theory of curvature elasticity that deals with small deviations of the nematic director from a uniformly and perfectly oriented nematic. This continuum theory is a first order elastic theory in what concerns bulk elasticity. OZF elasticity describes satisfactory conventional nematics and its drawbacks and failures have been widely discussed \cite{saupe,luiz}.
	Recently, a new class of achiral nematics  with a periodic structure in the nanoscale range has been identified \cite{cestari,panov,imrie} exciting great interest in the nematic liquid crystal
community. A few nematic modulated phases have been observed but their exact structure remains under investigation even for the most familiar among them, the twist-bend nematic $\mathrm{N_{TB}}$  phase \cite{Borshch,Zimmer,vana_exp,chen} that is also termed $\mathrm{N_{x}}$ nematic \cite{vana_exp}. Several models have been proposed for the $\mathrm{N_{TB}}$ \cite{dozov,shamid,virga,greco,ferrarini,pre,longa,kats,matsu,vana_th,lelidis,barbero}. In particular, models implying elasticity can be grouped to two categories those requiring a negative Frank elastic constant \cite{dozov,shamid,lelidis,barbero} and those that do not \cite{virga,pre,matsu,kats}. A softening of the bend elastic constant arises from its renormalization due to flexoelectricity \cite{shamid} and/or polar effects steaming from molecular shape \cite{vana_th,osipov}. That means, one has to introduce either a new element of symmetry like the helical axis unitary vector remaining in the frame of linear elasticity, or to expand the elastic energy to higher order. Nevertheless, elasticity is not the only way to obtain modulated nematics, for instance, biaxiality \cite{longa} and/or polar order \cite{vana_th}, or entropy \cite{ferrarini} are some other options. Therefore it seems that at the time being the understanding of modulated nematics remains poor and decisive experiments in order to qualify or disqualify some models are still lucking. According to Dozov's paper \cite{dozov}, the modulation arrives because the bend elastic constant becomes negative in $\mathrm{N_{TB}}$. This hypothesis implies that in order to describe a spontaneously deformed nematic phase of achiral molecules, OZF elastic theory has to be extended to include gradients of the deformation tensor. Another approach which is widely applied in classical elasticity of solids \cite{brugger,barsch,chang,russi,landau} consists in expanding $f=f(\nabla{\bf n})$ to higher powers than the second, of the deformation tensor, has been recently applied in nematics without further justification \cite{lelidis,barbero}. Hereafter, we refer to this approach as extended first order elasticity (E1OE).
	
	In the present paper, we extend the OZF continuum elastic theory by expanding the elastic free energy density, $f$, up to fourth order  (4OE) in the derivatives of $\mathbf{n}(\mathbf{r})$ so as to describe situations where one or more Frank's elastic constants become softer or negative, that is, $f=f(\nabla{\bf n},\dots,\nabla\nabla\nabla\nabla{\bf n})$. The tensor fields of the elastic constants are decomposed in their invariants. Moreover, the E1OE theory is derived in a systematic way from the invariants of the elastic constants. We apply both theories in the case of the $\mathrm{N_{TB}}$ and we compare their respective results. Finally, the splay bend nematic  $N_{SB}$ phase is investigated.

%%%%%%%%%%%%%%%%%%%%%%%%%%%%%%%%%%%%%%%%%%%%%%%%%%%%%%%%%
\section{Non-linear nematic elasticity}
%%%%%%%%%%%%%%%%%%%%%%%%%%%%%%%%%%%%%%%%%%%%%%%%%%%%%%%%%	
OZF elastic theory was obtained under the assumption that $f$ is an analytic function of the elastic tensor $\nabla{\bf n}$. Elastic deformations have to be mild at molecular scale. When the length-scale of the deformation becomes comparable to the molecular length, the linear elasticity fails. A possible generalization would involve spatial derivatives of ${\bf n}({\bf r})$ of higher order than the first. In order to operate such an expansion of $f$ one needs a criterion to quantify the relative importance of derivatives and of their powers that enter in each term of the expansion. Such a criterion can be provided by molecular models if the intermolecular interaction energy is known. Once one relates intermolecular interaction to the elastic constants \cite{supmat,saupe,luiz}, it results that the effective order of a term in the expansion of $f$ results from the sum of the order of all derivatives composing that term, for instance, the terms $(\mathrm{d}n/\mathrm{d}x)\,(\mathrm{d^{k-1}}n/\mathrm{d}x^{k-1})$ and $(\mathrm{d^{2}}n/\mathrm{d}x^{2})\,(\mathrm{d^{k-2}}n/\mathrm{d}x^{k-2})$ are both of order $k$.

%%%%%%%%%%%%%%%%%%%%%%%%%%%%%%%%%%%%%%%%%%%%%%%%%%%%%
%\section{Elastic energy}
%%%%%%%%%%%%%%%%%%%%%%%%%%%%%%%%%%%%%%%%%%%%%%%%%%%%%
Applying this rule, the bulk elastic energy density, of a uniaxial nematic composed by achiral molecules, up to fourth order terms is given by
	\begin{eqnarray}\label{elen}
	f &=& f_0+K_{ijkl}n_{i,j} n_{k,l}
	+ N_{ijk}\, n_{i,jk}\\
	&+&H_{ijklmnpq}\, n_{i,j} n_{k,l} n_{m,n} n_{p,q}+ G_{ijklmn}\, n_{i,jk} n_{l,mn}\nonumber\\
	& +& M_{ijklpqr}\,n_{i,j}n_{k,l}n_{p,qr}+ P_{ijklrs}\,n_{i,j}n_{k,lrs}+Q_{ijklr}\,n_{i,jklr}\nonumber
	\end{eqnarray}
where $f_0$ is the energy density of the state with uniform alignment (undeformed), and $i,j,k,\ell,r,s,p,q=x_1,x_2,x_3$. $K_{ijkl}$ and $ N_{ijk}$ are second order terms while the rest of terms in  (\ref{elen}) are fourth order terms. Linear and third order terms vanish identically due to the non-polar character of nematic phases, $f({\bf n})=f(-{\bf n})$, and are not presented here. Of course, these latter terms are present in chiral nematics \cite{saupe,berreman}. Using standard techniques for the calculation of the invariants of a tensor field, we calculated the invariants of the elastic tensors apperaring in (\ref{elen}). Hereafter, we use this decomposition in order to investigate the $N_{TB}$, and $N_{SB}$ for which there is some experimental evidence. These two phases represent unidimensional problems, that is, ${\bf n}$ depends  on just one spatial coordinate, say, ${\bf n}={\bf n}(x_3)$. Therefore only a few among over than $50$ invariants entering in $f$ survive and the subsequent analysis is simplified.
	
Using the condition ${\bf n}={\bf n}(x_3)$, we find three second order invariants
\begin{eqnarray}
\label{1}U_1=n_3^2(n_{1,3}^2+n_{2,3}^2+n_{3,3}^2),\quad U_2=n_{3,3}^2,\quad U_3=n_{1,3}^2+n_{2,3}^2+n_{3,3}^2
\end{eqnarray}
which can be rewritten as
$U_1=|{\bf b}|^2$, $U_2=s^2$, and $U_3=s^2+t^2+|{\bf b}|^2$
in terms of splay, twist, and bend deformations defined \cite{prost} by
$s=\nabla \cdot {\bf n}$, $t={\bf n}\cdot \nabla \times {\bf n}$,  and ${\bf b}={\bf n}\times \nabla \times {\bf n}
$. In the same representation, the six fourth order invariants can be written as
\begin{eqnarray}
\label{18}V_1=|{\bf b}|^4,\quad V_2=s^2 |{\bf b}|^2,\quad V_3= |{\bf b}|^2(s^2+t^2+ |{\bf b}|^2),\\
\label{21}V_4=s^4,\quad V_5=s^2(s^2+t^2+ |{\bf b}|^2),\quad V_6=(s^2+t^2+ |{\bf b}|^2)^2
\end{eqnarray}
Finally, the total elastic energy density can be written as
\begin{eqnarray}\label{1Delen}
f=f_0+f_2+f_4=f_0+\frac{1}{2}\,\sum_{i=1}^3 K_i\,U_i+\frac{1}{4}\,\sum_{i=1}^6 H_i\,V_i
\end{eqnarray}
where $f_2$ is related to the invariants of second order, and $f_4$ to those of fourth order. As can be easily verified, the fourth order contribution, $f_4$, is a homogeneous expression in $s^2$, $t^2$, and $|{\bf b}|^2$ as was supposed in \cite{barbero}.

%%%%%%%%%%%%%%%%%%%%%%%%%%%%%%%%%%%%%%%%%%%%%%%%%%%%%
\section{Extended first order elasticity}
%%%%%%%%%%%%%%%%%%%%%%%%%%%%%%%%%%%%%%%%%%%%%%%%%%%%%
%%%%%%%%%%%%%%%%%%%%%%%%%%%%%%%%%%%%%%%%%%%%%%%%%%%%%
\subsection{Twist-bend case}
%%%%%%%%%%%%%%%%%%%%%%%%%%%%%%%%%%%%%%%%%%%%%%%%%%%%%
First, we describe the approach of extended first order elasticity in the cases of $TB$ and $SB$ nematic.
To begin, let us apply the results of the above analysis to the heliconical twist-bend deformation characterized by the nematic director
\begin{equation}
\label{tw1}{\bf n}=(\cos \phi \,{\bf u_1}+\sin \phi\,{\bf u_2})\sin \theta_0+ {\bf u_3}\,\cos \theta_0
\end{equation}
where the conical angle $\theta_0$ is position independent and $\phi=\phi(x_3)$.
Using (\ref{1Delen}), the elastic energy density of the $N_{TB}$ phase is given by
\begin{equation}
\label{tw13}f=f_0+\frac{1}{2}R(\theta_0)\,\phi'^2+\frac{1}{4}S(\theta_0)\,\phi'^4
\end{equation}
where $\phi '=\mathrm{d}\phi/\mathrm{d} x_3$, and
\begin{eqnarray}
\label{tw11}R(\theta_0)&=&(K_1\,\cos^2 \theta_0+K_3)\,\sin^2 \theta_0,\\
\label{tw12}S(\theta_0)&=&(H_1\,\cos^4\theta_0+H_3\,\cos^2 \theta_0+H_6)\sin^4\theta_0
\end{eqnarray}
are effective elastic constants.
The Euler-Lagrange equation of (\ref{tw13}) is
\begin{equation}
\label{tw14} \left[R(\theta_0)+S(\theta_0)\phi'^2\right]\phi'=\alpha
\end{equation}
where $\alpha$ is an integration constant. For $R=S=0$ or $R\ge 0$ and $S>0$ only the uniform nematic solution exists. Modulated solutions may appear for $R<0$ and $S>0$. In the absence of surface anchoring energy, $\alpha=0$, from (\ref{tw14}) we obtain
\begin{equation}
\label{tw15}\phi'=\phi_u'=0,\quad \phi'=\phi'_d=\pm\sqrt{\frac{|R(\theta_0)|}{S(\theta_0)}}
\end{equation}
corresponding to a uniform, and spontaneously deformed states respectively.  Comparison of their energy implies that the deformed state is energetically favorable if permitted to exist, since
\begin{eqnarray}
\label{tw17}f(\phi_d')=f_0-\frac{(K_1 \cos^2 \theta_0+K_3)^2}{4(H_1\cos^4\theta_0+H_3\cos^2\theta_0+H_6)}<f_0=f(\phi_u')
\end{eqnarray}
The value of $\theta_0$ has to be determined by minimizing (\ref{tw17}).

Let us consider, as example, the case where $K_1<0$, $K_3>0$ and $H_6>0$. The energy of the deformed state is rewritten as
\begin{equation}
\label{tw19}f(\theta_0)=f_0-\eta \frac{(1-\xi \cos^2 \theta_0)^2}{1+h_3\, \cos^2\theta_0+h_1\,\cos^4\theta_0}
\end{equation} where, we introduced
\begin{equation}
\label{tw18} \xi=|K_1|/K_3,\quad h_1=H_1/H_6>0,\quad h_3=H_3/H_6<0,\quad{\rm and}\quad \eta=K_3^2/4H_6
\end{equation} Note that $K_3=K_{22}$ and $K_1=K_{33}-K_{22}$. Minimization gives the modulated solutions $\cos^2\theta_0=-(h_3+2\xi)/(2 h_1+h_3\xi)$ which correspond to the twist-bend phase. If one considers the case $K_{33}<0$ then  $\xi>1$, a standard analysis shows that the twist bend configuration is the ground state when  $1<h_1<\xi^2$ and $-2\sqrt{h_1}<h_3<-2(h_1+\xi)/(1+\xi)$.
Figure 1 shows $f(\theta_0)/\eta$, blue curve, and the effective elastic constants of second $R(\theta)$, orange curve, and fourth $S(\theta)$, green curve, order. $f_0$ is normalized to 0. The conical angle that minimizes the energy is $\theta_0=23.7^o$.
\begin{figure}[h]
	\centering
	\includegraphics[width=7cm]{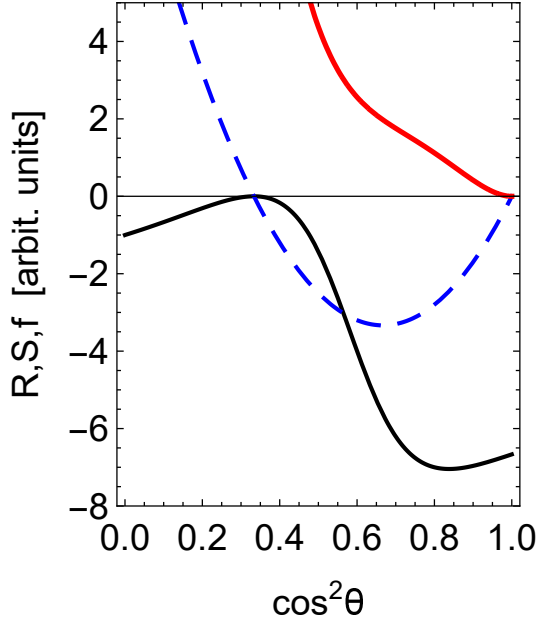}
	\caption[]{Energy and effective elastic constants vs $\cos^2\theta$ in reduced units $f/\eta$ (black solid line), $R/K_3$ (blue dashed line) and $S/H_6>0$ (red thick line). The curves are not in scale in the vertical direction in order to be visible. $f_0=0$.}
	\label{Figure_1}
\end{figure}
%%%%%%%%%%%%%%%%%%%%%%%%%%%%%%%%%%%%%%%%%%%%%%%%%%%%%
\subsection{Splay-bend case}
%%%%%%%%%%%%%%%%%%%%%%%%%%%%%%%%%%%%%%%%%%%%%%%%%%%%%

Let us consider now the splay-bend nematic phase which is also a one-dimensional deformation, always in the frame of the E1OE, and in the absence of anchoring energy. In this framework, indicating by $\theta$ the angle formed by the nematic director ${\bf n}$ with the $x_3$ axis, and assuming that ${\bf n}$ is contained in the $(x_1,x_3)$-plane then the nematic director components are $
n_1=\sin \theta$, $n_2=0$, $n_3=\cos \theta$. The invariants of second order are
\begin{eqnarray}
\label{s2}U_1=\cos^2 \theta\,\theta'^2,\quad U_2=\sin^2\theta\,\theta'^2,\quad U_3=\theta'^2
\end{eqnarray}
and those of fourth order are
\begin{eqnarray}
\label{s6}V_1=\cos^4\theta\,\,\theta'^4,\quad V_2=\sin^2\theta\,\cos^2\theta\,\,\theta'^4,\quad V_3=\cos^2\theta\,\,\theta'^4\\
\label{s9}V_4=\sin^4\theta\,\,\theta'^4,\quad V_5=\sin^2\theta\,\,\theta'^4,\quad V_6=\theta'^4
\end{eqnarray}
The elastic energy density is cast in the form
\begin{equation}
\label{s12}f=f_0+\frac{1}{2}{\cal R}(\theta)\theta'^2+\frac{1}{4}{\cal S}(\theta)\theta'^4
\end{equation}
with the effective elastic constants
\begin{eqnarray}
\label{s13}{\cal R}(\theta)&=&K_1\,\cos^2\theta+K_2\,\sin^2\theta+K_3,\\
\label{s14}{\cal S}(\theta)&=&H_1\,\cos^4\theta+H_2\,\sin^2\theta\,\,\cos^2\theta+H_3\,\cos^2\theta+
H_4\,\sin^4\theta+H_5\,\sin^2\theta+H_6
\end{eqnarray}
The Euler-Lagrange equation of the problem is
\begin{equation}
\label{5-s}4[{\cal R}(\theta)+3{\cal S}S(\theta)\theta'^2]\theta''+[2 {\cal R}(\theta)+3 {\cal S}(\theta) \theta'^2]\theta'^2=0
\end{equation}
The absence of interaction between the substrate and the liquid crystal is mathematically responsible for the transversality conditions
\begin{equation}
\label{6-s}[{\cal R}(\theta) + {\cal S}(\theta) \theta'^2]\theta'=0,\quad{\rm at}\quad x_3=\pm d/2
\end{equation}

%%%%%%%%%%%%%%%%%%%%%%%%%%%%%%%%%%%%%%%%%%%%%%%%%%%%%%%%%

Let us consider first the simple case where ${\cal R}(\theta)=-\kappa$ and ${\cal S}(\theta)=H$, with $\kappa$ and $H$ independent of $\theta$, and  $\kappa>0$. In this situation the total elastic energy density is
\begin{equation}
\label{7-s}f=f_0-\frac{1}{2}\kappa \theta'^2+\frac{1}{4}H \theta'^4
\end{equation}
with solutions $\theta'= 0$ and $\theta'=\sqrt{\kappa/ H}$.
The stable solution is  the deformed one since
\begin{equation}
\label{11-s}f(\theta'=0)=f_0,\quad{\rm and}\quad f\left(\theta'=\sqrt{\frac{\kappa}{H}}\right)=f_0-\frac{1}{4}\frac{\kappa^2}{H}
\end{equation}
In this particular case, the tilt angle is a monotonic function of $x_3$. This conclusion can be generalized. Suppose that in a given point $x_{30}$ along the $x_3$ axis,  $\theta'(x_{30})=0$.  (\ref{5-s}) implies
\begin{equation}
\label{12-s}{\cal R}[\theta(x_{30})]\theta''(x_{30})=0.
\end{equation}
Since ${\cal R}\neq 0$, it follows that $\theta''(x_{30})=0$. Similarly,  one can show that $\theta'''(x_{30})=0$ and so on. Hence, either $\theta'=0$, that is $\theta$ is position independent, or $\theta'$ cannot change sign, that is $\theta(x_3)$ is a monotonic function. Nevertheless, a deformed state that minimizes the energy implies ${\cal R}(\theta) <0$, and hence $\theta$ cannot be position independent. Therefore, in the framework of the E1OE, we infer that the tilt angle of the nematic director for the $\mathrm{N_{SB}}$ phase is a monotonic function of the position  $\theta=\theta(x_3)$.

%%%%%%%%%%%%%%%%%%%%%%%%%%%%%%%%%%%%%%%%%%%%%%%%%%%%%%%%%
\section{Fourth order elasticity}
%%%%%%%%%%%%%%%%%%%%%%%%%%%%%%%%%%%%%%%%%%%%%%%%%%%%%%%%%
In the approximation of fourth order elasticity $f$ depends on derivatives of $\mathbf{n}(\mathbf{r})$ up to fourth order. For twist-bend deformation, substitution of $\mathbf{n}(\mathbf{r})$ from (\ref{tw1}) into the expressions of the invariants results to the elastic energy
\begin{equation}
\label{h1}f=\frac{1}{2}R\phi'^2+\frac{1}{4}S \phi'^4+\frac{1}{2}G \phi''^2+H \phi' \phi'''
\end{equation} where $R$, $S$, $G$, and $H$ depend on $\theta_0$.
The last term can be decomposed into a bulk term that renormalizes the elastic constant $G$, and to a surface-like term since $\phi' \phi'''=(\phi' \phi'')'-\phi''^2$. Therefore, disregarding the third derivative term results to neglect surface-like terms in the energy $F$.

%%%%%%%%%%%%%%%%%%%%%%%%%%%%%%%%%%%%%%%%%%%%%%%%%%%%%%%%%
\subsection{In presence of surface-like terms}
%%%%%%%%%%%%%%%%%%%%%%%%%%%%%%%%%%%%%%%%%%%%%%%%%%%%%%%%%
In the absence of surface-like terms, $f$ reduces to
\begin{equation}
\label{g1}f=f_0+\frac{1}{2}R \phi'^2+\frac{1}{4} S \phi'^4 +\frac{1}{2} G \phi''^2
\end{equation}
 For $R(\theta_0)\ge 0$, the undeformed solution is the stable one. Hereafter, we suppose that $R(\theta_0)<0$. For a sample in the form of a slab of thickness $d$, and the $x_3-$axis perpendicular to the bounding surfaces at $x_3=\pm d/2$,  the minimization of the total energy
\begin{equation}
\label{h0}F=\int_{-d/2}^{d/2} f(\phi',\phi'')\,\mathrm{d}x_3
\end{equation}in the absence of surface anchoring energy, gives the first integral
\begin{equation}
\label{g6}R \phi'+S \phi'^3-G \phi'''=0
\end{equation}
with the  boundary conditions $
\label{g7} \phi''=0$ at $x_3=\pm d/2$.
Apart from the trivial uniform nematic solution $\phi_u'=0$, Eq.(\ref{g6}) has a second solution
\begin{equation}
\label{g11}\phi_d'=q_d=\sqrt{-\frac{R}{S}}=const.,
\end{equation}
to which corresponds the position independent elastic energy density
\begin{equation}
\label{g12}f_d=f_0-\frac{R^2}{4 S}<f_u=f_0
\end{equation}
Note that if the quartic term in $\phi'$ is neglected, no deformed solution exists.
Finally, a third solution, $\phi'=q(x_3)$, with variable wave-vector exists. However, $f(q(x_3))>f(q_d)$, as can be shown by substituting $q(x_3)=q_d+\delta q(x_3)$ in the energy
\begin{equation}
f(q_d+\delta q(x_3))-f(q_d)=\frac{1}{4}S\,\delta q^2 (2 q_d+\delta q)^2 +\frac{1}{2} G\,\delta q'^2>0,
\end{equation}
and therefore $q_d$ corresponds to the absolute minimum of (\ref{g1}).

%%%%%%%%%%%%%%%%%%%%%%%%%%%%%%%%%%%%%%%%%%%%%%%%%%%%%%%%%
\subsection{In presence of surface-like terms}
%%%%%%%%%%%%%%%%%%%%%%%%%%%%%%%%%%%%%%%%%%%%%%%%%%%%%%%%%

In the following, we investigate the full energy density expression (\ref{h1}), that is, keeping surface-like terms. Minimizing
\begin{equation}
F=\int_{-d/2}^{d/2} f(\phi',\phi'',\phi''')\,dx_3,
\end{equation}
we get the first integral
\begin{equation}
\label{h7}R q+S q^3+(2 H-G)q''=0,
\end{equation}
where $q=\phi'$, and the boundary conditions
\begin{eqnarray}
\label{h8}(G-H)q'= 0\quad\&\quad H q = 0\quad \text{at}\quad x_3=\pm d/2
\end{eqnarray}
Since the ordinary differential equation (\ref{h7}) is of second order and the boundary conditions to be satisfied are four, the function minimizing the total energy will be in general discontinuous. One encounters a similar problem as the well known $K_{13}$ puzzling, for some time, question \cite{oldano,durand,vertogen}.

\begin{figure}[h]
	\centering
	\includegraphics[width=8cm]{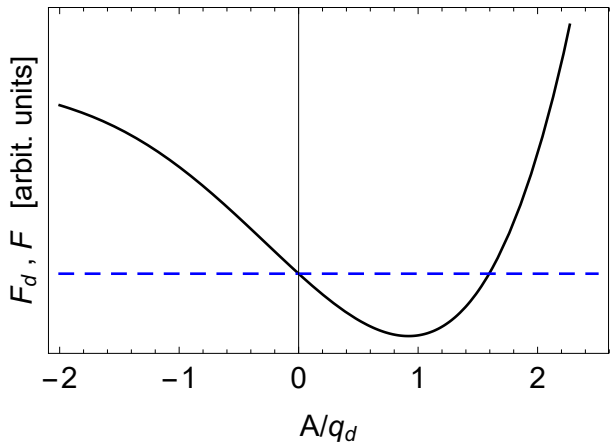}
	\caption[]{Energy with surface like terms $F$ (black solid line) and position independent deformation energy $F_d$ (blue dashed line)  vs $A$ for a nematic slab of thickness $d$. $F$ has a minimum,  $F_{min}<F_d$, for $A\ne 0$. $f_0=0$.}
	\label{Figure_2}
\end{figure}

A trivial solution  of (\ref{h7}) is $q=0$, and the corresponding energy density is $f=f_0$. However, a simple inspection shows that the functions varying rapidly enough near to the limiting surfaces are related to lower total energy. As an example let us consider the trial function
\begin{equation}
\label{h14}q(z)=q_d+A\,\frac{\sinh(x_3/L)}{\sinh(d/2L)},
\end{equation}
defined in $0\leq x_3 \leq d/2$, and continued analytically in even manner in $-d/2\leq x_3\leq 0$. In (\ref{h14}),  $A$ is a constant, $q_d$ is given by (\ref{g11}), and $L\ll d$ is the thickness of a surface layer. This trial function coincides with $q_d$ in the bulk, and differs from it just in the surface layers. The total energy of the sample is a function of the amplitude $A$. A plot $F=F(A)$ shows that $F-F_0<0$ reaches a minimum for $A\neq 0$ (see Figure 2). A direct calculation of the profile can be performed by minimization of the total energy if $q(x_3)$ is expanded as a power series of $x_3-x_3^*$, with $x_3^*=d-L$ in the surface layers. In this framework, it is assumed that
\begin{eqnarray}
\label{h16} q &=& q_d,\quad {\rm for}\quad0\leq x_3\leq x_3^*\\
\label{h17} q &=& q_d+b (x_3-x_3^*)+(c/2)(x_3-x_3^*)^2, \quad{\rm for}\quad x_3^*\leq x_3\leq d/2
\end{eqnarray}
where $b$ and $c$ are free parameters. Evaluating $F$, see Figure 3, and minimizing it with respect to $b$ and $c$ shows that the minimizing function is discontinuous with discontinuity points at the border. For both signs of $H$, $F$ presents a minimum.
Therefore, the stable solution in the bulk is $q_d$ and hence for the twist-bend phase $q=\sqrt{-R/S}$, as has been determined in the case of the E1OE model. Note that, in the present case, it seems that the $(\phi')^4$ term controls the bulk solution over the $(\phi'')^2$ term.

\begin{figure}[h]
	\centering
	\includegraphics[width=8cm]{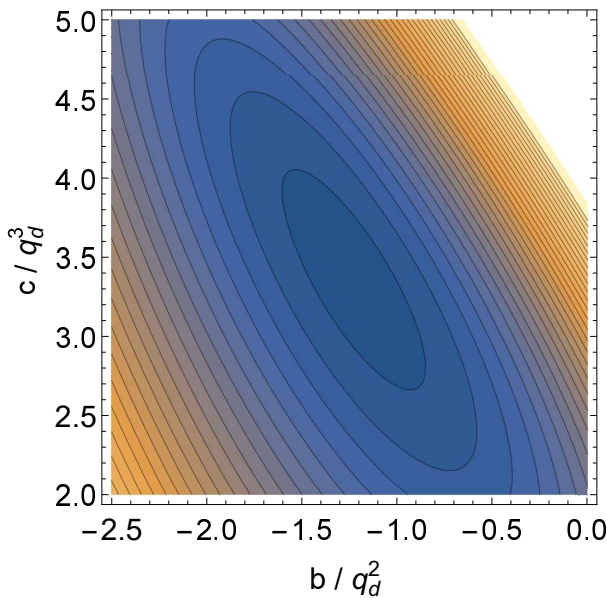}
	\caption[]{Contour plot of $F$ vs $b$ and $c$. $F$ has a minimum for $b/q_d^2 =-1.22777$, and $c/q_d^3 = 3.36218$. $f_0=0$.}
	\label{Figure_3}
\end{figure}

\section{Conclusions}
In conclusion, we extended in a systematic way the linear elasticity of nematics using two approaches. First by taking into account gradients of the nematic strain tensor, and second by considering higher powers of the deformation tensor. Applying both models in the case of a twist-bend nematic we found that the bulk solution is the same. In the case of the strain tensor gradient model, surface effects arise in the same way as in the problem of $K_{13}$ of undeformed nematics. In the case of a splay-bend nematic, in the framework of the extended first order elasticity, we demonstrated that small oscillations of the nematic director $\mathbf{n}(x_3)$ around the $x_3-$axis are forbidden, and $\mathbf{n}(x_3)$ is a monotonic function of $x_3$. Finally, we note that the tight pitch ($\approx 10\mathrm{nm}$) of the helix in the nematic twist-bend phase sheds some doubts about the applicability of a continuum theory. For a detailed discussion on this issue, see for instance \cite{vana_th}. Nevertheless, this phase has been first predicted by continuum models which at least qualitatively seem to describe the up to now known physics of the phase. Further, a pitch of $\sim10\mathrm{nm}$ was predicted by the elastic model \cite{pre} as shown in \cite{rosseto}. Certainly the modulated nematic phases problem is far to be elucidated.

%%%%%%%%%%%%%%%%%%%%%%%%%%%%%%%%%%%%%%%%%%%%%%%%%%%%%
\section*{Appendix: Elasticity from molecular interactions}
%%%%%%%%%%%%%%%%%%%%%%%%%%%%%%%%%%%%%%%%%%%%%%%%%%%%%

The molecular approach is based on molecular interactions which are supposed to be additives and to decrease rapidly with separation so that can be neglected for a length much longer than a molecular dimension. In a nematic liquid crystal, the anisotropy of the intermolecular interaction gives rise to  anisotropic elastic constants.

We assume a uniaxial nematic composed by rod like molecules and with perfect nematic order $S=1$, that is,  the molecular long axes coincide with the nematic director. Let ${\bf n}={\bf n}({\bf R})$ and  ${\bf n'}={\bf n}({\bf R'})$ be the directors of two interacting molecules at the points ${\bf R}$ and ${\bf R'}={\bf R}+{\bf r}$. The two body interaction energy between two molecules is a function of their relative orientation and their separation  \cite{saupe}
\begin{equation}
\label{f5}v=v({\bf n},{\bf n'},{\bf r})
\end{equation}
The interaction energy between two elements of volume $d \tau$ and $d\tau'$ at ${\bf R}$ and ${\bf R'}$ containing $dN=\rho({\bf R}) d\tau$ and $dN'=\rho({\bf R'}) d\tau'$, where $\rho$ is the particle density, is
\begin{equation}
\label{f5-1}d^2 {\cal V}=v d\tau d\tau'.
\end{equation}
Supposing constant density of particles, assumption valid just in the bulk, (\ref{f5-1}) can be rewritten as \cite{saupe}
\begin{equation}
\label{f52}d^2 {\cal V}=g({\bf n},{\bf n'},{\bf r}) d\tau d\tau'
\end{equation}
where $g({\bf n},{\bf n'},{\bf r})=\rho^2 v ({\bf n},{\bf n'},{\bf r})$.
In the elastic approximation $v\neq 0$ only for $r_m\leq r\leq r_M$, where $r_m$ is a lower cut-off, and $r_M$ is of the order of the range of the molecular forces responsible for the condensed phase. If ${\bf n}$ varies slowly over $r_M$
we have
\begin{equation}
\label{f6}{\bf n'}={\bf n}({\bf R'})={\bf n}({\bf R})+\delta {\bf n}({\bf R},{\bf r}),
\end{equation}
with $|\delta {\bf n}({\bf R},{\bf r})|\ll 1$. Hereafter, we limit our analysis to second order. However, the results can be generalized to all orders. Substituting (\ref{f6}) into the expression for $g$ we get
\begin{equation}
\label{f7}g=g({\bf n},{\bf n}+\delta {\bf n},{\bf r}),
\end{equation}
 Since $|\delta {\bf n}({\bf R},{\bf r})|\ll 1$ for $r_m\leq r\leq r_M$, we can expand $g$ in powers of $\delta {\bf n}$
\begin{equation}
\label{f8}g=g({\bf n},{\bf n},{\bf r})+q_i \delta n_i+\frac{1}{2} q_{ij} \delta n_i \delta n_j+{\cal O}(3),
\end{equation}
where Einstein summation convention for repeated indexes is assumed, and
\begin{equation}
\label{f9}q_i({\bf n},{\bf r})=\left(\frac{\partial g}{\partial n_i'}\right)_{{\bf n'}={\bf n}},\quad{\rm and}\quad q_{ij}({\bf n},{\bf r})=\left(\frac{\partial^2 g}{\partial n_i' \partial n_j'}\right)_{{\bf n'}={\bf n}}
\end{equation}
 To derive the elastic energy density we expand $\delta n_i$ in power series of $r_i$, the cartesian components of ${\bf r}$
\begin{equation}
\label{f10}\delta n_i=n_{i,j} r_j+\frac{1}{2}n_{i,jk} r_j r_k+.....
\end{equation}
where $n_{i,j}=(\partial n_i/\partial X_j)$ are evaluated at ${\bf R}$. Substituting (\ref{f10}) into (\ref{f8}) results to
\begin{equation}
\label{f11}g=g({\bf n},{\bf n},{\bf r})+q_i n_{i,j} r_j+\frac{1}{2}\left(q_i n_{i,kl}+q_{ij} n_{i,k} n_{j,l}\right) r_k r_l+...
\end{equation}
The elastic energy density, in the mean field approximation, at ${\bf R}$, is given by
\begin{equation}
\label{f12}f=\frac{1}{2}\int\int\int_{\tau} g({\bf n},{\bf n'},{\bf r}) d\tau',
\end{equation}
where $\tau$ is the volume of the sample. Due to the short range character of the interaction the integral has to be performed on a volume of linear dimension of the order of $r_M$.
Substituting (\ref{f11}) into (\ref{f12}) we get
\begin{equation}
\label{f13}f=f_0+{\cal L}(\nabla{\bf n})+{\cal N}\nabla(\nabla{\bf n})+\frac{1}{2}{\cal K}(\nabla{\bf n})(\nabla{\bf n})
\end{equation}
where the elements of the tensors ${\cal L}$, ${\cal N}$ and ${\cal K}$ are
\begin{eqnarray}
\label{f14}L_{ik}&=&\frac{1}{2}\int\int\int_{\tau} q_i r_k \,d\tau',\\
\label{f15}N_{ikn}&=&\frac{1}{4}\int\int\int_{\tau}q_i r_k r_n \,d\tau',\\
\label{f16}K_{ijkn}&=&\frac{1}{2}\int\int\int_{\tau} q_{ij}r_k r_n \,d\tau'.
\end{eqnarray}
Expansion (\ref{f11}) gives a rule to expand the elastic energy density to spatial derivatives higher than the first order. For instance, $n_{i,j} n_{k,l}$ is of second order as $n_{j,kl}$. $n_{i,jkl}$ is of third order as $n_{i,j} n_{k,l} n_{p,q}$ and $n_{i,jk} n_{r,s}$, and so on.

\end{document}